\documentclass[5p,time]{elsarticle}

\usepackage[spaces,hyphens]{url}
\usepackage{lineno,hyperref}
\modulolinenumbers[5]
\hypersetup{pdfauthor={Name}}

\usepackage{graphicx}
\usepackage{titlesec}
\usepackage{multirow}
\usepackage{calrsfs}
\usepackage{tikz} % for latex diagram
\usepackage{textcase}
\usepackage[tablename=TABLE]{caption}
\DeclareCaptionTextFormat{up}{\MakeTextUppercase{#1}}
\usepackage[T1]{fontenc}% optional T1 font encoding
\usepackage{graphicx}
\usepackage{amsmath}
\usepackage{amssymb,amsfonts}

\usepackage{balance}
\usepackage{longtable}
\usepackage{amssymb}
\usepackage{array}
\usepackage{pgfplots}
\usepackage{tikz}
\usetikzlibrary{arrows,shapes,positioning,shadows,trees}
\usepackage{forest}
\usetikzlibrary{arrows.meta}
\tikzstyle{arrow} = [thick,->,>=stealth]
\usepackage{longtable}
\pgfplotsset{compat=1.14}
\usepackage{multicol}
\definecolor{lava}{rgb}{0.81, 0.06, 0.13}
\usepackage[normalem]{ulem}

\usepackage{supertabular}

\journal{Elsevier}

\begin{document}

\begin{frontmatter}

\title{A Review on Edge Analytics: Issues, Challenges, Opportunities, Promises, Future Directions, and Applications}

\author{Sabuzima Nayak}
\ead{sabuzma\_rs@cse.nits.ac.in}
\author{Ripon Patgiri\fnref{myfootnote}}
\ead[URL]{http://cs.nits.ac.in/rp/}
\ead{ripon@cse.nits.ac.in}
\author{Lilapati Waikhom}
\ead{lilapati\_rs@cse.nits.ac.in}
\address{National Institute of Technology Silchar, India}
\fntext[myfootnote]{Corresponding author: Ripon Patgiri, Department of Computer Science \& Engineering, National Institute of Technology Silchar, Cachar-788010, India}

\author{Arif Ahmed}
\ead{arif.ahmed@ericsson.com}
\address{Ericsson, Sweden}

% \author[mymainaddress,mysecondaryaddress]{Elsevier Inc}

% \author[mysecondaryaddress]{Global Customer Service\corref{mycorrespondingauthor}}
% \cortext[mycorrespondingauthor]{Corresponding author}

% \address[mymainaddress]{1600 John F Kennedy Boulevard, Philadelphia}
% \address[mysecondaryaddress]{360 Park Avenue South, New York}

\begin{abstract}
Edge technology aims to bring Cloud resources (specifically, the compute, storage, and network) to the closed proximity of the Edge devices, i.e., smart devices where the data are produced and consumed. Embedding computing and application in Edge devices lead to emerging of two new concepts in Edge technology, namely, Edge computing and Edge analytics. Edge analytics uses some techniques or algorithms to analyze the data generated by the Edge devices. With the emerging of Edge analytics, the Edge devices have become a complete set. Currently, Edge analytics is unable to provide full support for the execution of the analytic techniques. The Edge devices cannot execute advanced and sophisticated analytic algorithms following various constraints such as limited power supply, small memory size, limited resources, etc. This article aims to provide a detailed discussion on Edge analytics. A clear explanation to distinguish between the three concepts of Edge technology, namely, Edge devices, Edge computing, and Edge analytics, along with their issues. Furthermore, the article discusses the implementation of Edge analytics to solve many problems in various areas such as retail, agriculture, industry, and healthcare. In addition, the research papers of the state-of-the-art edge analytics are rigorously reviewed in this article to explore the existing issues, emerging challenges, research opportunities and their directions, and applications. 
\end{abstract}


\end{frontmatter}

% \linenumbers

\section{Introduction}
Cloud Computing gives a sense of infinity to the technology, i.e., huge storage, high computational resources, extensive infrastructure, and so on \cite{CC}. These are the features of Cloud Computing that attracted the attention of everyone. User can expand their system capability anytime with a few extra costs. In smart technology, all devices are smart to support the fast pace of the world. With the support of the Cloud, smart computation becomes easy \cite{Khan}. However, the increase in smart devices leads to the issue of Big data \cite{Xiong}. Cisco estimated that around 2025, the number of smart devices would reach 500 billion \cite{camhi}. Vast volumes of data where most of the data are unwanted were transferred to the Cloud for computation. Thus, the bandwidth and network resources are used to transfer unwanted data, which the Cloud eventually ignores or deletes. Moreover, privacy became another important issue of Cloud Computing because it transfers personal data to a far place. Moreover, another important issue of the Cloud is its long distance from smart devices. Transmitting the data to the Cloud, computing the data, and again transmitting the user's output response increased the latency. All these issues lead to the requirement of performing computation near smart devices. Thus, another technology emerged called Edge technology. Edge computing is an emerging technology to meet the growing demands of users and scale up with the ever growing network accesses. Cloud Servers are far from the users and their devices; therefore, Satyanarayanan \textit{et al.} \cite{Satyanarayanan} proposed to bring the Cloudlet near to the smart devices. Thus, Edge Computing emerges, and it becomes the future backbone of all smart devices, for instance, Unmanned Aerial Vehicles.

In Edge technology, the smart devices are also called Edge devices, i.e., smartwatches, smartphones, routers, switches, mobile access devices, and IoT gateways \cite{Safavat}. The Edge devices are made capable of collecting, processing, and analyzing the user data and returning the response quickly. Therefore, these devices demand proper security measurement \cite{Sha}. The computation performed by the Edge devices is called Edge computing, and the analysis performed by the Edge devices is called Edge analytics. Edge technology aims to provide features of Cloud Computing in Edge devices. However, Cloud has everything, and logically, it is nearly infinite, i.e., storage, processor, infrastructure, etc., but Edge devices have everything limited. Edge devices have limited power, limited storage, limited resources, and processor with low computational capability \cite{Pan}. The compute-intensive application requires enormous resources, powerful processors, huge memory, and high power. However, Edge devices are small devices; hence, embedding compute-intensive applications is impossible because many smart devices are carried or worn by people, such as smartphones and smartwatches. Nonetheless, Edge technology has solved many issues and brought Cloud Computing nearer to the edge. 
%An Edge device has all these in small size, hence, within such restrictions supporting a compute-intensive application is difficult. 

Edge analytics execute some applications on the Edge devices to analyze the data. It made the Edge devices capable of responding to small requests very quickly. For example, in an industry, all machines have sensors to measure some parameters. If these parameters cross an upper or lower threshold, it may damage the product or machine. Thus, it is essential to observe such parameters constantly. But, these parameters cannot be observed efficiently by humans, and in a few cases, it may be dangerous. Edge devices can monitor these parameters in such situations and take some predefined action before the problem goes out of control. 

\begin{table}[!ht]
\caption{Review papers published on Edge Analytics}\footnotesize
\begin{tabular}{|p{1cm}|p{1cm}|p{2.5cm}|p{1.5cm}|}
    \hline
\textbf{Paper} & \textbf{Year} & \textbf{Paper Name} & \textbf{Topic Covered} \\ \hline \hline

Sharma \textit{et. al.} \cite{Sharma} & 2018, IEEE &  Edge Analytics for Building Automation Systems: A Review & Smart building \\ \hline

Zhang \textit{et. al.} \cite{Zhang1} & 2019, IEEE & Edge Video Analytics for Public Safety: A Review & Public Security, Video Edge analytics  \\ \hline

Djamal-uddin \textit{et. al.} \cite{djamaluddin}& 2019, Society of Petroleum Engineers & Towards Real-Time Edge Analytics-A Survey Literature Review of Real-Time Data Acquisition Evolution in the Upstream Oil and Gas Industry & \vspace{0.5cm}Industry \\ \hline

    \end{tabular}
    \label{review}
\end{table}

Table \ref{review} lists the review papers published on Edge analytics. However, these papers have reviewed Edge analytics in a specific domain. In our article, we cover broader areas of state-of-the-art Edge Analytics. We uncover issues and challenges associated with Edge analytics. Also, it provides in-depth insight into applications of Edge analytics. Initially, Section \ref{EA} provides a short discussion on Edge analytics. We classify the Edge analytics applications based on the type of data in Section \ref{TEA}. Moreover, machine learning and artificial intelligence are also explored for embedding in Edge devices; however, all these algorithms are compute-intensive. Hence, Edge analytics requires fewer computational and low memory-consuming algorithms. This important point is elaborated, and the reviews of various techniques are presented in section \ref{MEA}. Edge analytics is deployed in many areas such as healthcare, transportation, etc. Section \ref{sol} reviews some Edge analytics techniques to illustrate how Edge analytics provides solutions to issues present in these areas. Moreover, Edge technology makes the devices smart, which is further elaborated in section \ref{ST}. Section \ref{Dis} presents a brief discussion on Edge Analytics issues and some other aspects. Finally, Section \ref{Con} concludes the article.

\section{Overview on Edge Analytics}
\label{EA}

Edge analytics is a promising technology that is able to create many new possibilities and opportunities. It is the most emerging technology in technology and it will dominate the entire market in the near future \cite{Gartner}. Edge analytics perform analysis on data collected by Edge devices \cite{GigaSight,Cao,Nikolaou}. Therefore, it is a merger of Artificial Intelligence and IoT \cite{Jin}. After collecting the data, Edge analytics execute some analytics algorithm, for example, video analysis techniques or machine learning algorithms. After the analysis, Edge analytics initiates some predefined action such as notifying the authority or raising the alarm. The computation performed in an Edge device is called Edge computing. Also, Edge analytics requires computing. Thus, Edge analytics is a part of Edge computing. 

\begin{figure}[!ht]
    \centering
    \scalebox{0.7}{
    \begin{tikzpicture}
        \begin{scope}[opacity=1]%blend group=hard light]
            \fill[red!30!white]   ( 90:1.2) circle (4);
            \fill[green!30!white] (90:1) circle (3);
            \fill[blue!30!white]  (90:3) circle (2);
        \end{scope}
        \node at ( 90:-2.4)    {Edge Devices};
        \node at (90:0.1)    {Edge Computing};
        \node at (90:3)    {Edge Analytics};
        % \node [font=\Large] {\LaTeX};
    \end{tikzpicture}
    }
    \caption{Relationship among Edge devices, Edge Computing and Edge analytics using Venn diagram}
    \label{EA-venn}
\end{figure}
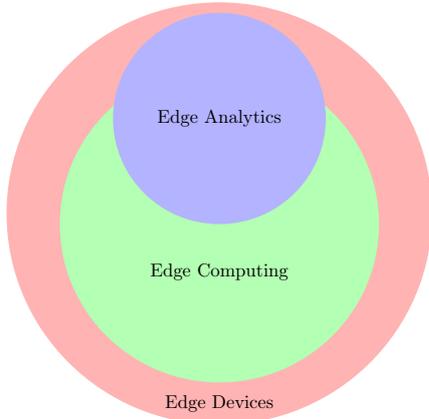

In figure \ref{EA-venn}, the Edge technology is illustrated using a Venn diagram. Currently, Edge technology has three aspects, namely, Edge devices, Edge computing, and Edge analytics. An Edge device has two components, namely, hardware and software. Figure \ref{EA-venn} only considers the software component of Edge devices. In software, Edge computing refers to computation performs on edge devices, such as IoT devices \cite{EKhan}. Edge analytics deploy single or many applications for data analysis. These applications also perform computation, i.e., Edge computing. Hence, the larger part of Edge analytics is Edge computing. Rest small part of Edge analytics is for data transformation, output data formatting, etc. Data transformation is to modify the data as per the format, for example, partitioning whole video frames into subsets of frames for video Edge analytics. Similarly, the Edge analytic output needs to be modified, filtered, or compressed to reduce file size for storage in Cloud. Another point to include is fog computing. Fog computing is a layer present between Edge and Cloud \cite{Martinez, Dong}. Cisco defined fog computing as “an extension of the cloud computing paradigm that provides computation, storage, and networking services between end devices and traditional cloud servers” \cite{bonomi}. We distinguish among Edge, Fog, and Cloud. However, a discussion on fog computing is excluded in this article. 

Some domains such as healthcare or industries constantly observe some parameters because in case these parameters cross a lower or higher threshold may lead to disastrous consequences. For example, a few parameters measured or observed in industries are pressure, temperature, etc. Sometimes, these parameters may cross a threshold value affecting the quality of the product, damaging the product/machine, causing accidents (e.g., fire or explosion), endangering the workers or surrounding civilization due to unknown reasons. These may incur huge losses or even closer to the industry. However, observing such parameters is not sufficient; timely action is also essential. Human evaluation may not always be possible; thus, Edge analytics is most appropriate in such scenarios. 

The Edge devices such as sensors continuously collect the parameter readings. The upper or lower limits are recorded in the Edge devices. Based on those threshold values, the Edge analytics in Edge devices execute some analytics algorithm to determine whether the parameters have crossed the threshold value or not. In case the parameters have crossed the threshold value, appropriate action is taken, such as reducing or shutting down the source that produces the parameter (e.g., shut down fire supply to reduce pressure) or inform a human to evaluate the condition. These actions are also predefined in Edge analytics. Moreover, Edge analytics sends filtered data to Cloud for further or future analysis. The industries have many different machines, each with different parameters. Observing all machines using Edge devices is very convenient. However, the role of the Cloud cannot be eliminated. Edge devices have limited storage; hence, data are transmitted to Cloud for storage. In addition, some analysis requires the execution of high computational algorithms such as artificial intelligence (AI) algorithms. In such scenarios, the training based on data is performed in the Cloud. Then the Cloud sends the produced inferences to the Edge devices. Based on the inferences, the Edge analytics test the data. In this way, the Edge devices are made capable of using complex algorithms in Edge devices. Currently, executing complex algorithms in Edge devices is explored. This will help to reduce the dependence of Edge devices on the Cloud, specifically in computation and analysis. Figure \ref{EA-dig} illustrates the various Edge devices connected to the Edge gateway and provides data analysis either from Edge analytic or Big Data analytics. Big Data analytics is present in Cloud.

\subsection{Edge Analytics and Big Data analytics}
\begin{figure}[!ht]
    \centering
    \includegraphics[width=0.4\textwidth]{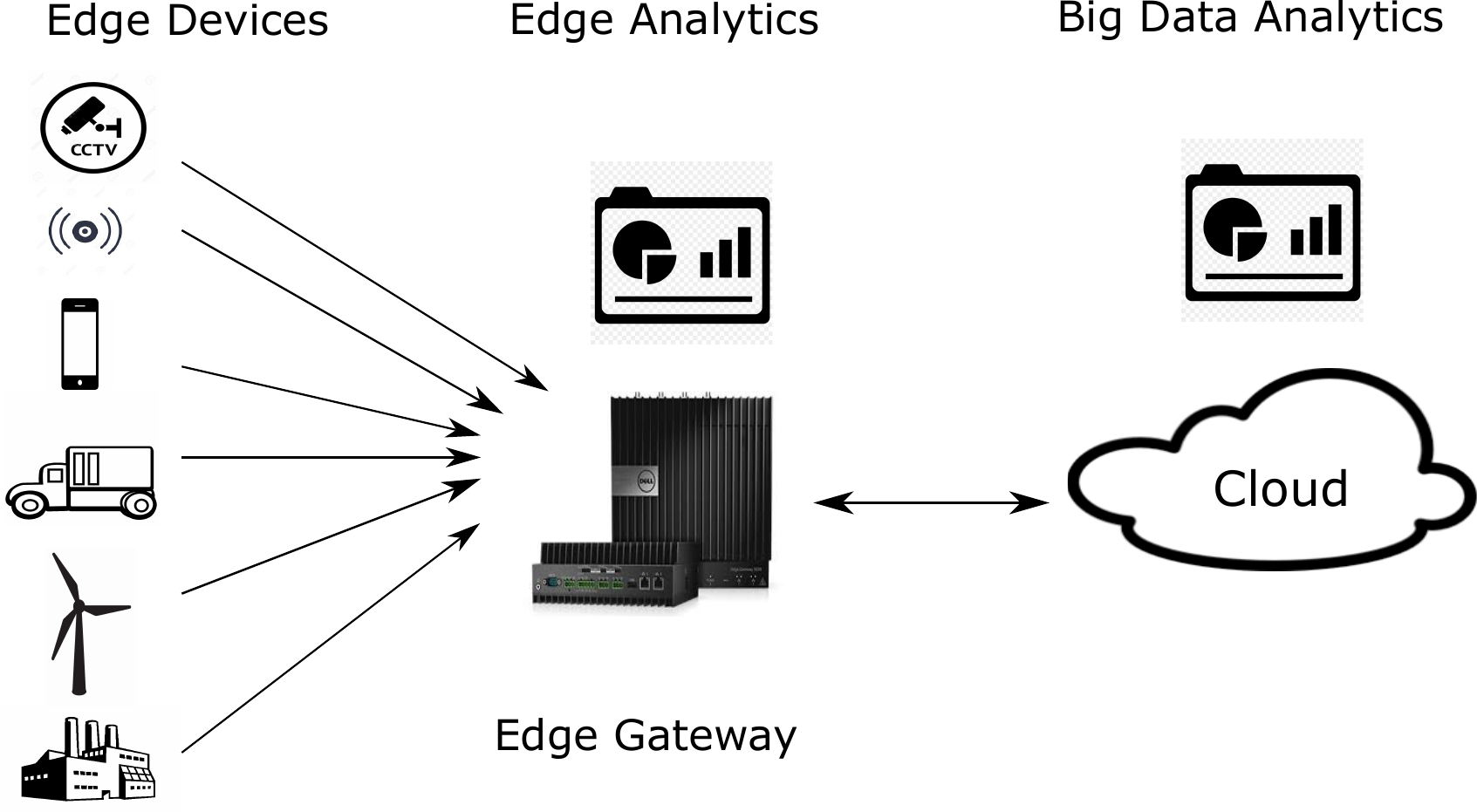}
    \caption{\textbf{Relation between Edge Analytics and Big Data Analytics}}
    \label{EA-dig}
\end{figure}

Big Data analytics is appropriate in situations where huge volumes of data are collected, and those data have uncertain value, patterns, or unknown outcomes. Big Data analytics requires the support of large numbers of servers, storage devices, infrastructure, and distributed systems. In addition, it requires advanced analytics and Big computing system to support advanced and sophisticated technologies to compute the information. Big Data analytics are present at Cloud because Cloud is capable of providing the huge resources required by Big Data analytics. However, Edge technology also wants to deploy data analytics in Edge devices. Because, in that case, the source of the data can itself perform the analysis. However, Edge Analytics is in the development stage. Edge technology is trying to fit an Elephant (Big Computing system) in a small box (i.e., router). It will reduce the communication cost, bandwidth, along with other benefits. However, the Edge analytics applications are still not advanced nor sophisticated. Currently, the power supply to support such advanced applications is also an issue. Therefore, Edge analytics is dependent on the Cloud in case of a request that requires high computing. Table~\ref{BA_tab} represents the difference between Big Data analytics and Edge analytics. Figure \ref{EA-dig} shows the hierarchy of implementation of the Edge analytics and Big Data analytics.
 
\begin{table}[!ht]
    \centering
    \caption{\textbf{Big Data Analytics vs Edge Analytics}} \footnotesize
    \begin{tabular}{|p{2cm}|p{2cm}|p{2cm}|}
      \hline \centering
         \textbf{Parameters}& \textbf{Big Data Analytics} & \textbf{Edge Analytics}  \\ \hline \hline
         Deployed & Cloud & Edge Device (e.g., Routers, Sensors)\\ \hline
         Age & Older & Younger \\ \hline
         Development & Matured & Developing \\ \hline
         Size of data & Big data & Small \\ \hline
         Application & Advanced and sophisticated & Not advanced and unsophisticated \\ \hline
         Storage & Distributed System & Small memory of Edge device \\ \hline
         QoS dependence & None & Cloud \\ \hline 
    \end{tabular}
    \label{BA_tab}
\end{table}

\subsection{Design and Deployment Requirements}
Edge analytics is performed on image, video, and data collected from Edge devices such as CCTV, mobile, and sensor. A few examples of the applications are reading license plates, monitoring crowd density or river levels. The Edge nodes need to perform analysis with limited power and computation. Thus, some design and deployment considerations are required for obtaining good results by Edge analytics \cite{GSMA}. 
\begin{itemize}
    \item Data quality: The data should be good enough to recognize the surrounding. Because in some cases, the surrounding data is more relevant than the main object. For example, searching for a car with a specific number in an image taken by a CCTV is on the road. 
    \item Data format: Data can be of various formats. Edge analytics data need to be converted to an appropriate format; for example, a video may require to divide into a specific number of frames. Thus, an open format for data is required for implementing Edge analytics in a wide range of data. 
    \item Data analytics: Some Edge analytics application requires to have a clear parameter definition. These specific parameters help in monitoring. Moreover, in the case of deploying machine learning algorithms, the definition of the topography in the image or landscape of audio needs to be defined to train the algorithm.
    \item Edge device setup: In some cases, fixed Edge devices help improve the processing, such as CCTV cameras. The fixed position helps in fixed reference points, which gives more accurate results. 
\end{itemize}

Currently, the Edge devices have many constraints. Hence, all computation cannot be performed on Edge devices. The Edge devices handle some simpler and less compute-intensive tasks. Thus, partitioning of computation between Edge devices and Cloud is also a requirement. Planner \cite{planner} is a middleware for cost-efficient, transparent, and uniform stream analytics on Edge and Cloud. It effectively and automatically partitions the computations between Edge and Cloud. This amalgamation is called hybrid stream processing which improves the system performance by increasing flexibility. The data is filtered and collected locally, making the processing of live-stream data very easy using the hybrid method. Planner follows two models, namely, the resources model and the network cost model. The resource model is used for stream processing. A network cost model is used for communication and streamflow between Edge and Cloud resources. Various network and compute resources are represented as a directed graph for computation (stream graph). It helps in the partition of computation between Cloud and Edge analytics. 

\section{Types of Edge Analytics}
\label{TEA}
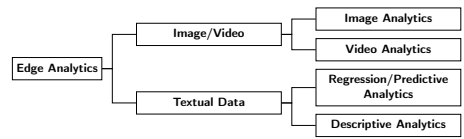
\begin{figure}[!ht]
\centering
\scalebox{0.45}{
\newcolumntype{C}[1]{>{\centering}p{#1}}
\begin{forest}
 for tree={
  		if level=0{align=center}{% allow multi-line text and set alignment
    		align={@{}C{40mm}@{}},
  		},
  		grow=east,
  		draw,
  		font=\sffamily\bfseries,
  		edge path={
    		\noexpand\path [draw, \forestoption{edge}] (!u.parent anchor) -- +(3mm,0) |- (.child anchor)\forestoption{edge label};
  		},
  		parent anchor=east,
  		child anchor=west,
  		l sep=10mm,
  		%tier/.wrap pgfmath arg={tier #1}{level()},
  		%edge={ultra thick, rounded corners=2pt},
  		%fill=white,
  		%rounded corners=2pt,
  		%drop shadow,
	}
  [Edge Analytics
    [Textual Data
        [Descriptive Analytics]
    	[Regression/Predictive Analytics]
    ]
    [Image/Video      
      [Video Analytics]
      [Image Analytics]
    ]
  ]
\end{forest}
}
\caption{\textbf{Types of Edge Analytics}}
\label{EAtype}
\end{figure}

The Edge analytics is classified based on the type of data collected by the Edge devices, namely, textual or image/video. Based on the video, Edge analytics is of two types: image analytics and video analytics. Similarly, based on textual, Edge analytics is of two types: regression/predictive analytics and descriptive analytics. 

\subsection{Edge Analytics based on Image/Video}
Based on the video data set, the Edge analytics is classified into image Edge analytics and video Edge analytics. Image Edge analytics has a video data source because the application takes a single frame of video, which is also an image as input. Whereas video Edge analytics takes a video as input. Due to less computing power and resource, the video processing algorithm is not executed on a whole video; rather, a video is divided into sets of video frames \cite{ROI,Bailas}. In the rest of the section, a short review of various Edge analytics techniques is presented. Moreover, some important features and remarks related to these techniques are presented in Table \ref{VEA_tab}.

% privacy
GigaSight \cite{GigaSight} is a decentralized hybrid Cloud architecture that implements video Edge analytics. It is a virtual machine (VM)-based Cloud. In the architecture, the Edge node is called cloudlets which performs the video analysis. Cloudlet performs video stream modification for the preservation of privacy. Cloudlets store videos for a short period based on billing policies. Each video is denatured in a personal VM in the cloudlet. Denaturing is the automatic lowering of fidelity in a video stream based on user choice for preserving privacy. Besides the Edge device of the owner, this VM only has access to the original video to maintain privacy. The denaturing process is executed on a single set of video frames at a time because it is very compute-intensive when applied to the whole video. After denaturing, two files are obtained as output, namely, thumbnail and encrypted video stored in different cloudlets. The thumbnail video gives an overview of video content that is used for indexing and query operations. Binary filtering based on metadata (e.g., location and time) is performed on these files to make the video frame entirely blank or pass without modification. This filtering has less computational complexity. Then, content-based filtering is performed. The throughput of the indexing operation depends on the number of objects that need to be detected. But, the number of objects is high; hence, the classifier is applied to the most important objects in the database. Moreover, due to this reason, indexing is performed in another VM. During the search operation, two hierarchical searchings are performed for finding a video based on user-specific context. The first search is traditional SQL searching which is performed using the metadata and tags obtained from indexing. The second search is performed using actual content, which gives more relevant videos to the user. But, this search is performed in parallel because it is highly compute-intensive. GigaSight also performs time-sensitive video searching. The result after video processing and its metadata are saved in the Cloud. 

%Denaturing balances between privacy and value. 
ROI aided video encoder system \cite{ROI} is proposed to detect objects of interest using object detection and tracking framework. The framework used is Faster-RCNN \cite{R-CNN} where RCNN is a region-based convolutional neural network. The system has three components: object detector-tracker framework, video Edge analytics engine, and the H.264 encoder. The frames of the video are separated into different groups based on importance. The importance is determined using the position of the objects of interest. The video Edge analytics engine receives a group of frames and marks Macro Blocks (MB) based on importance. It uses the VGG-16 network for the extraction of features to detect the object of interest. Object detection is expensive; hence, the task is computed in a single node called an aggregator node. Each Edge node sends one JPEG encoded frame to the aggregator node and executes the object tracker on the rest $N-1$ frames locally because the tracker process has low complexity. However, executing the object detection algorithm only in the aggregator node reduces the distributed computation. H.264 encoder optimizes rate distortion to reduce the quantization parameter to reduce cost. 
% Transfer of frames to aggregator node does not create much overhead on bandwidth usage. Increase in successive frame interval increases the number of unexplained MBs. 

Bailas \textit{et al.} \cite{Bailas} proposed video Edge analytics for crowd monitoring. The Edge analytics uses metadata for crowd counting. Some metadata are density, crowd activity, or the number of people gathered. The video frames are converted to grayscale, and the background is extracted. The computation is performed on important objects in the video. To determine the crowd density, first, the frame is divided into grids, and the percentage of grid cells occupied by the people is calculated. This approach executes object recognition algorithm proposed by He \textit{et al.} \cite{He}. The algorithm uses deep learning and has average time complexity. In addition, it has good precision on low-quality images. For crowd counting, the algorithm proposed by Marsden \textit{et al.} \cite{Marsden} is implemented. It uses heat maps to calculate the number of people. However, this algorithm requires high computation and gives bad precision in high quality video.

Xie \textit{et al.} \cite{Xie1} proposed a video Edge analytics that give centimeter-level positioning accuracy with low latency response. The architecture has four layers, namely, application, controller \& orchestration, Edge computing, and video capture. The application layer is responsible for providing computer vision algorithms, security monitoring, area management, and augmented reality. The controller layer provides detailed specifications to video Edge analytics. Orchestration allocates resources. The architecture implements container technology for lightweight virtualization and partitions the video Edge analytics process into modules. Lightweight virtualization prevents service migration issues such as environmental configuration and compatibility. The result obtained from a module that is common to multiple video Edge analytics is shared. A scheduling algorithm is used to initiate the execution of a module at the appropriate time using fewer resources. The parameters such as video resolution and video sampling rate are defined based on user requirements. Based on the parameters, the controller determines the video Edge analytics application. Similarly, based on configurations, the video Edge analytics process is divided into subtasks which are executed in different modules. The orchestration provides the resources, and the scheduling algorithm executes the modules.
%The result obtained by a module is forwarded to the next module. Each module maintains a different namespace.

Jang \textit{et al.} \cite{Jang} proposed a video Edge analytics that adjusts its configuration dynamically. The architecture has three main components: Edge device (camera), controller, and knowledge registry. The physical camera uses virtualization to have multiple virtual cameras to execute more number of video Edge analytics applications. The hypervisor monitors video frames to determine the context changes. However, it reduces the QoS of the application. Hypervisor sends the configuration setting and video frames to application containers. The hypervisor has four components: video streamer, context monitor, quality estimator, and configuration manager. The video streamer transmits the video frames from the physical to the virtual Edge camera. The context monitor updates the context change periodically. The configuration manager modifies the virtual camera configuration based on the required QoS. The controller consists of three components: cluster resource manager, camera selector, and executor. The cluster resource manager manages the Edge cameras and their context values. It also maintains the network information to determine the context changes in the Edge camera. The camera selector selects the appropriate Edge camera. The executor delivers a processed message to the Edge camera. Knowledge registry has both service logic and service ontology databases. Service ontology contains the conceptual relationships among camera configuration, application requirements, and contexts. The user makes a video request (VR) using a message which contains VR name, service type, application requirements, and filtering parameters. The controller uses service type for the selection of the target Edge camera. Every IoT camera calculates its service capability, which the controller uses to determine the target camera. After receiving the VR, the controller refers to the knowledge registry for determining the application-aware camera configuration. There is an interpreter that translates the high-level to application-specific requirements using a rule engine. The rule engine executes the respective rules. Application-specific requirements are further translated into camera-level requirements based on the relationship defined in application logic. The application logic establishes the relationship that maps specific requirements to configuration.

%video Edge analytics
Chen \textit{et al.} \cite{Chen1} proposed a video Edge analytics embedded with indoor recognition, face recognition and semantic analysis. In the video Edge analytics, each application is separated into different units. Virtualization is used to handle complexity and resource allocation issues. The Edge device architecture consists of a video capture layer, Edge computing layer, controller \& orchestration layer, and application layer. The controller and orchestration layer has two modules, namely, network controller and container orchestration. The network controller manages the network resources and the Edge network environment. Container orchestration maintains the resource status of each Edge node and schedules resources to the Edge analytics unit. The video Edge analytics runs in the application layer. The live video is given as input to video Edge analytics, then the indoor recognition and face recognition application is executed. Semantic analysis is performed on the video files saved in the video library. The container orchestration monitors the resources in case resources are not available, then the services with relaxed restrictions are paused to make the resources available to important tasks. The face recognition application is executed in every video frame, where the frame is taken every second. It leads to the consumption of more resources. 
%Moreover, the system should filter the video frames to reduce computation.

%privacy
$REVAMP^2T$ (real-time edge video analytics for multicamera privacy-aware pedestrian tracking) \cite{Neff} is a video Edge analytics that supports privacy. It detects, reidentifies, and tracks the person across many cameras without storing the live streaming. The infrastructure maximizes hardware utilization in Edge analytics. The $REVAMP^2T$ algorithm pipeline has three main phases: detection, reidentification and, tracking \& prediction. The video frames received from the cameras are given as input to the detection network. The detections generated as output are preprocessed, such as scaling to required size and aspect ratio. Then detections are forwarded to the feature extractor to generate a feature vector for each detection. Furthermore, spatial filtering is performed in the local database in parallel. OpenPose \cite{Cao1} is used for detection. It predicts pose using part affinity fields to process image input and gives an output of person detections with marked keypoint locations. The key points yield the person’s body motion which is used for action recognition and motion prediction. The reidentification algorithm produces discriminative features for detection comparison and local database matching. $REVAMP^2T$ implement MobileNet-V2 \cite{Jacob} which is a lightweight deep convolution network. MobileNet-V2 creates a feature extraction network for extracting discriminative features from each detection. After completion of the reidentification process, the current detections are transferred to Long-short-term-memory (LSTM) network. The detection is matched with the next five frames to rectify misdetection (in case), which may have been caused due to fewer resources of Edge analytics. 
\begin{table*}[!ht]
    \centering
    \caption{Video Edge Analytics techniques} 
    \begin{tabular}{|p{1.8cm}|p{8.5cm}|p{6cm}|}
      \hline 
         \textbf{Techniques} & \centering \textbf{Features} & \hspace{0.8cm} \textbf{Remarks}  \\ \hline \hline
	
GigaSight \cite{GigaSight} & \small{$\bullet$ Video stream modification preserves privacy \newline $\bullet$ Thumbnail video helps in indexing and query operations \newline $\bullet$ Metadata based filtering has less computational complexity} & \small{$\bullet$ Denaturing process is too compute-intensive \newline $\bullet$ Searching process is compute-intensive \newline $\bullet$ Video is deleted periodically based on cost policies \newline $\bullet$ Indexing operation has high time complexity} \\ \hline

ROI aided video encoder system \cite{ROI} & \small{$\bullet$ Tracker process has low complexity \newline $\bullet$ H.264 encoder optimises rate distortion \newline $\bullet$ Optimised rate distortion reduces the quantization parameter \newline $\bullet$ Optimised rate distortion reduces cost \newline $\bullet$ Executing object detection algorithm only in aggregator node reduces computational burden on other Edge nodes \newline $\bullet$ Frame transfer to aggregator node has less bandwidth overhead} & \small{$\bullet$ Increase in successive frame interval increases number of unexplained macro blocks \newline $\bullet$ Executing object detection algorithm only in aggregator node reduces the utility of distributed system features} \\ \hline

Ballas \textit{et al.} \cite{Bailas} & \small{$\bullet$ Metadata is used for crowd counting \newline $\bullet$ Provides better data privacy compared to image \newline $\bullet$ Object recognition algorithm has an average time complexity \newline $\bullet$ Object recognition algorithm has good precision on low quality images} & \small{$\bullet$ Crowd counting algorithm requires high computation \newline $\bullet$ Crowd counting algorithm gives bad precision in high quality video \newline $\bullet$ Increase in number of videos decreases the accuracy and precision} \\ \hline

Xie \textit{et al.} \cite{Xie1} & \small{$\bullet$ Positioning accuracy up to centimeter-level with low latency response \newline $\bullet$ Implements container technology for lightweight virtualization \newline $\bullet$ Lightweight virtualization prevents service migration issues \newline $\bullet$ Sharing result of the common module reduces processing time and saves resources} & \small{$\bullet$ High resolution process takes more resources \newline $\bullet$ Execution of high resolution process interfere other tasks by consuming more resources} \\ \hline

Jang \textit{et al.} \cite{Jang} & \small{$\bullet$ Adjust the configuration dynamically \newline $\bullet$ Virtualization helps to execute more number of video analytics applications \newline $\bullet$ VR messages sometimes contain multiple VRs for better selection of Edge cameras \newline $\bullet$ Cluster Resource Manager maintains the network information which helps in determining the context changes} & \small{$\bullet$ Monitoring of video frames by hypervisor reduces the QoS \newline $\bullet$ Reduction in dimming duration leads to an increase in CPU overhead difference \newline $\bullet$ Sub-components in each component increases the complexity of the technique} \\ \hline

Chen \textit{et al.} \cite{Chen1} & \small{$\bullet$ Edge analytics is embedded with multiple video applications \newline $\bullet$ Maintains resources status of each task \newline $\bullet$ Resources are made available by pausing less important task(s) \newline $\bullet$ Provides accurate and fast video analytics services \newline $\bullet$ Virtualisation handles complexity and resource allocation issues} & \small{$\bullet$ Resource starvation \newline $\bullet$ Quality of face databases influences the accuracy of recognition \newline $\bullet$ Video application execute on video frame every second and consumes more resources \newline $\bullet$ Video frames are not filtered} \\ \hline

REVAM$P^2T$ \cite{Neff} & \small{$\bullet$ Image data are not stored or transmitted in network \newline $\bullet$ Reidentification algorithm differentiate between people without providing personal identification \newline $\bullet$ Error-friendly operations in MobileNet-V2 minimizes the error caused by the half-precision operations \newline $\bullet$ During communication failure Edge device buffer permits a level of data synchronization \newline $\bullet$ Edge nodes do not fight over available sockets \newline $\bullet$ Communication does not take resources from consistent reidentification} & \small{$\bullet$ Misdetection may occur \newline $\bullet$ Reidentification process is designed for short temporal window} \\ \hline 

\end{tabular}
\label{VEA_tab}
\end{table*}
% \end{tabular}
%   
% \end{table*}

\subsection{Edge Analytics based on Textual Data}
The predictive Edge analytics compute machine learning and AI algorithms to prepare a model. This model is used to predict the output of future data. In some cases, the model is computed in the Cloud, and only the inferences are sent to Edge devices. Those inferences are used by Edge analytics to analyze the data. This helps decrease the computation in Edge devices; however, it increases the dependence of Edge analytics on the Cloud. Descriptive Edge Analytics provides a description of the data set. In the rest of the section, the review of various data Edge analytics techniques is included with Table \ref{DEA_tab} listing these techniques advantages and remarks. 

%predictive analytics
Nikolaou \textit{et al.} \cite{Nikolaou} proposed predictive Edge analytics for data filtering in Edge nodes. The approach automatically determines the transmission of data from Edge nodes to a data center or Cloud. Data filtering is based on the sliding window. When the collected data crosses a window threshold, a regression model is trained upon the collected data. This gives a prediction on the cached model. The data is stored in the memory. The error between new and old predictions is calculated to determine whether to change the cached model or not. When the error crosses a threshold value, then new predictions are considered. Instead of data, cache model parameters are transmitted to the data center or Cloud to save bandwidth and reduce communication overhead. However, training of regression model takes more time for a big sized window. 

%descriptive analytics
Cao \textit{et al.} \cite{Cao} proposed a descriptive Edge analytics for mobile Edge nodes. Descriptive Edge analytics analyses and predict patterns from real-time data. It optimally balances the highly dynamic data and usage of resources at every level of the network. It helps to achieve scalability and low latency. However, the Edge analytic is complex and requires more power. The approach separated the data processing into various descriptive analytical processes to decrease the Edge node's computation overhead and bottlenecks. The vehicle sends GPS location coordinates every five seconds. When the Edge node receives the data, it determines whether the vehicle is mobile or not \cite{HUANG201910}. The mobility is determined using the distance the vehicle has covered between two consecutive periods. In case the distance covered is less than 15 meters, then the vehicle is considered stationary. The next process is data collection and determining other parameters by performing calculations on received data. For example, calculating the distance covered in a single trip using the GPS location and time taken. All these data are transmitted to Cloud or datacenter once every week for further analysis. 

% water networks
Babazadeh \cite{arduino1} proposed an anomaly detection system in water networks using data Edge analytics. It uses the LoRa platform. All tasks are performed sequentially in the Edge device, and threads are used to perform the tasks. Mainly two threads are used, called main and secondary. The main thread collects and filters the data, then performs data compression using a lossless data compression algorithm. The Edge analytics evaluate the compression rate to detect abnormalities or outliers. Then it is transmitted to the center. The secondary thread executes a callback function that handles the concurrence among the huge number of fast and slow tasks. This is an interrupt service routine. Two types of packets are used for data communication, namely, system settings and worst-anomaly. System setting packets consist of few bytes which are sent to the Edge devices. The worst anomaly packet is the packet carrying notification about the worst anomaly to the center.  The worst anomaly is those data that have the smallest value for filtered compression rate. This technique has a low communication cost and increases the lifetime of the battery-powered Edge devices. Nonetheless, it also has a few remarks like packet size limitation specified by LoRa, and latency is more due to sequential execution of tasks.

%data analytics system
APPOXimate computing in IoT (APPROXIOT) \cite{APPROXIOT} is a data Edge analytics algorithm that aims at reducing latency and optimizing the usage of computing resources. The algorithm is based on approximate computing and hierarchical processing. The Edge nodes situated over a wide range collectively construct a logical stream processing pipeline similar to a tree. This tree is considered as APPROXIOT. The main goals of APPROXIOT are resource efficiency, adaptability, and transparency. After receiving a client query request, APPROXIOT generates an approximate query response within a rigorous error range. However, it only supports approximate linear queries. APPROXIOT algorithm is an online sampling parallel technique. APPROXIOT does not coordinate with other Edge nodes to increase parallel computation and scalability. Sampling optimizes resource usage. Moreover, the degree of sampling is changed with respect to the change in available resources. APPROXIOT performs two types of sampling based on the type of data stream source: stratified \cite{Stratified} and reservoir \cite{Reservoir}. Stratified sampling is suitable for the data stream received from a known source. On the contrary, reservoir sampling is suitable for the data stream received from an unknown source. Stratified sampling decreases sampling error and increases sample precision. However, it requires the statistics knowledge of all sub-streams. But, obtaining this knowledge is impractical. Reservoir sampling is more resource-efficient. However, it may corrupt the statistical quality of the sampled data in case data streams belong to different sources. APPROXIOT algorithm compromises on performance and accuracy. 
% Moreover, adjustment of sampling parameters are not automatic. Increase in window size increases the latency.  It selects a subset of data where data subset size depends on the available resources in the Edge node. APPROXIOT is completely transparent. Analyst does not have to monitor the resources or change any code for current data analytics application/query.  

%privacy
EdgeSanitizer \cite{Xu1} is a mobile data Edge analytics with a deep inference framework. It achieves local differential privacy (LDP) by adding noise to learned features. Differential privacy is a technique for preserving strong and guaranteed privacy by adding random noise to raw user data \cite{Liu}. EdgeSanitizer has three main steps, namely, data minimization, data obfuscation, and data reconstruction. Data minimization minimizes the data volume. The data size is minimized by extracting features using deep learning, which helps in reducing communication costs. Moreover, the dimension of the high dimensional mobile data is reduced using an autoencoder. Data obfuscation is a privacy preserving step.  Random Laplacian noise is added to the features to have LDP. In the final step, i.e., data reconstruction, the features are rebuilt, and the data is transferred to the Cloud. 

%predictive analytics
Harth \textit{et al.} \cite{Harth2} proposed a predictive analytics with efficient aggregation. The mechanism is lightweight and distributed. Sensing and Actuator Nodes (SAN) are Edge devices that use local prediction to eliminate contextual data and sends the filtered data to the Edge node. In the Edge nodes, a bounded-loss approximation mechanism is implemented for data filtering. However, the mechanism has less analytics accuracy caused due to approximation/reconstruction. The predictive analytics are deployed in both SAN and Edge nodes. SAN collects the data and constructs a contextual data vector. The contextual data vector is then taken as input, and predictive analytics is executed to return the response to users. SAN calculates a local prediction error using the selective data delivery principle. The contextual data vector is sent to the Edge node when the error is above a threshold value. However, using an optimal prediction error threshold is important because a high threshold value increases communication overhead. However, it decreases the reconstruction possibility in the Edge node. On the contrary, a low threshold value increases information loss. The power supply in SAN is low compared to Edge nodes. Therefore, the prediction function has less complexity and requires less computation. In the absence of a data vector, the Edge node reconstructs the data vector to perform further processing, although it is an overhead. Three policies are proposed for the reconstruction of data context vectors. In the first policy, reconstructed context vectors are the first element of the sliding window. In the second policy, the vectors are calculated as the average of the vectors currently present in the window. In the third policy, the exponential smoothing algorithm determines the vectors. 

\begin{table*}[!ht]
    \centering
    \caption{Textual Data Edge Analytics techniques}
    \begin{tabular}{|p{1.5cm}|p{6.5cm}|p{6.5cm}|}
      \hline \centering
         \textbf{Techniques} & \hspace{1cm}\textbf{Features} & \hspace{1cm} \textbf{Remarks}  \\ \hline \hline

Nikolaou \textit{et al.} \cite{Nikolaou} & \small{$\bullet$ Saves bandwidth \newline $\bullet$ Filters unwanted data \newline $\bullet$ Reduces communication overhead} & \small{$\bullet$ Regression model training takes more time for big sized window \newline $\bullet$ Small sized window gives less accurate results for regression model training} \\ \hline
%\newline $\bullet$  Requires a less compute-intensive and low power consuming regression algorithm 

Cao \textit{et al.} \cite{Cao} & \small{$\bullet$ Optimally balances the highly dynamic data and resources usage at every level of network \newline $\bullet$ Scalable \newline $\bullet$ Low latency \newline $\bullet$ Separation of data processing into various descriptive analytical process decreases computation overhead on Edge node} & \small{$\bullet$ Descriptive analytics is complex \newline $\bullet$ Consumes high power} \\ \hline

Babazadeh \cite{arduino1}  & $\bullet$ Low communication cost \newline $\bullet$ Enhances lifetime of the battery-powered Edge device \newline $\bullet$ Filtered and compressed data are send to the data center & $\bullet$ Packet size is limited \newline $\bullet$ More latency due to sequential implementation \newline $\bullet$ Tasks are computed sequentially \\ \hline
 
APPROXIOT \cite{APPROXIOT} & \small{$\bullet$ Achieves approximate computing \newline $\bullet$ Efficient usage of resources \newline $\bullet$ Adaptable \newline $\bullet$ Transparent \newline $\bullet$ Sampling optimises the resource usage \newline $\bullet$ Stratified sampling decreases sampling error \newline $\bullet$ Reservoir sampling is more resource-efficient \newline $\bullet$ Scalable} & \small{$\bullet$ Compromises between performance and accuracy \newline $\bullet$ Stratified sampling requires statistics knowledge of all sub-streams where obtaining it is impractical \newline $\bullet$ Reservoir sampling corrupt statistical quality of the sampled data in case data streams belong to different sources \newline $\bullet$ Adjustment of sampling parameters are not automatic \newline $\bullet$ Increase in window size increases the latency \newline $\bullet$ Only supports approximate linear queries} \\ \hline

EdgeSani- tizer \cite{Xu1} & \small{$\bullet$ Noise is added to learned features to achieve local differential privacy \newline $\bullet$ Lightweight feature extraction process \newline $\bullet$ Features are mutual-independent} & \small{$\bullet$ Data minimization mechanism result in a negligible reconstruction error \newline $\bullet$ Privacy budget in LDP for balancing utility and privacy is higher than DP \newline $\bullet$ More noise addition leads to faster degradation rate of inferences} \\ \hline

Harth \textit{et al.} \cite{Harth2} & \small{$\bullet$ SAN sends filtered data to Edge node to reduce the memory consumption of Edge node \newline $\bullet$ Filtering by SAN reduces network traffic \newline $\bullet$ Prediction function has less complexity \newline $\bullet$ Prediction function requires less computation \newline $\bullet$ Increase in prediction error threshold decreases the reconstruction possibility in Edge node} & \small{$\bullet$ Reconstruction in Edge nodes is an overhead \newline $\bullet$ Approximation/reconstruction decreases the analytics accuracy \newline $\bullet$ Increase in prediction error threshold increases communication overhead \newline $\bullet$ Low prediction error threshold increases information loss} \\ \hline

  \end{tabular}
  \label{DEA_tab}
\end{table*}
% \end{longtable}

\section{Machine Learning and Edge Analytics}
\label{MEA}
Researchers are always interested in developing less compute-intensive machine learning algorithms. With this intention, Edge technology is an opportunity to embed AI and machine learning algorithms in Edge devices. It helps Edge nodes to learn from the available data and experience to generate actions for the events without being programmed. Inference takes less computing resources to execute and generate output for new data. Therefore, inference helps to respond to the user request with low latency when executed in Edge nodes. Microsoft proposed Azure IoT Edge \cite{Azure} to integrate Edge technology and machine learning. In AWS Greengrass \cite{Greengrass} machine learning inference is executed locally in Edge devices using cloud-based models.

The Edge nodes face diverse issues due to machine learning algorithms. Some of the issues are listed below \cite{aral}:
\begin{itemize}
\item Subset data: An Edge node receives a subset of data. Accordingly, the learning performed by the machine learning algorithm is not optimal. 
\item Dynamic data: Real-time data are dynamic. Hence, training performed by the machine learning algorithm becomes obsolete after some time. 
\item Compute-intensive: All machine learning algorithms are compute-intensive. Hence, Edge nodes with low in all aspects, i.e., power, computation, resources, etc., are unable to support such algorithms.
\end{itemize}

In the following part of the section, some techniques that try to combine machine learning and Edge analytics are reviewed. Moreover, Table \ref{ML_tab} provides the advantages and remarks of these techniques.  

Edge-deployed Convolution Neural Network (ECNN) \cite{song} is a distributed data analytics framework in Smart Grids. ECNN is a distributed CNN that enhances the performance of data aggregation and analytics. Many initial layers are a series of convolution layers along with filters, pooling, fully connected layers (FC), and implements Softmax function for data classification. Cloud and Edge devices have many layers of convolution layers and pooling. In addition, the Edge nodes perform classification using FC layers. Local data in these nodes generates local inference, and more data-intensive inference is generated at Cloud. The size of local inference is small, which reduces transmission cost. ECNN performs data aggregation, which helps in cross-area data inference and making a decision. 

%predictive analytics
Sarabia-Jacome \textit{et al.} \cite{Sarabia} proposed an Edge gateway architecture having a predictive analytics module that implements deep learning models. The architecture is secure and maintains privacy. Moreover, architecture is interoperable. Limited resources are managed by using the container-based virtualization technique. Predictive analytics determines and predicts the patterns in the data received from Edge devices. This is achieved using services based on a deep learning model whose training is executed in the Cloud. The generation of inference using deep neural networks is high compute-intensive and consumes huge resources. Hence, the inferences are generated in the Cloud. Then, the inferences are sent back to Edge devices. The Edge devices have a tri-axial accelerometer sensor that records the data of the user.  The sensor data are in the form of tri-axial data $(x,y,z)$. The data is preprocessed using a sliding window which is later forwarded to the predictive analytics module. This module stores the inference generated in the Cloud and uses them to make decisions based on an event. The results of the event are stored in a temporal database of the Edge device. 
%Container-based technique uses the resources in a homogeneous manner. 

Moon \textit{et al.} \cite{Moon} proposed a framework for data analysis by distributing analysis roles across Edge analytics in various Edge nodes. This framework utilizes the Cloud resources optimally and also uses Edge nodes for immediate computation. Edge nodes collect the data and also monitor and control the actuators directly. In Cloud, machine learning algorithms are executed to generate models. These models are transmitted to Edge analytics based on the collected data by the Edge devices. All data are not transmitted to Cloud for data privacy. Hence, Cloud only accesses the open data. The Cloud consists of two modules, namely, IoT Data Management and Machine Learning Model Management, whereas Edge nodes have an IoT Interaction module. The first module collects and processes the data for training. The second module trains the data. The IoT Interaction module generates predictions based on the user profile. This framework is more dependent on the Cloud. The models in the Edge analytics become obsolete after some time and lead to incorrect decisions.

%CNN, DCNN, 
Dey and Mukherjee \cite{dl} proposed Edge analytics based on the deep learning framework. The framework provides schemes for resource distribution in Edge devices constrained using a model based on the Deep Convolution Neural Network (DCNN). However, tuning of the hyperparamters of the DCNN result in significant usage of resources. Edge analytics follows three steps for resources management, namely, resource standardization of intermediate resources, intermediate resource provisioning, and offload execution to support tiny devices \cite{Ai}. The advantages of standardizing resources are low latency and data manageability. Moreover, it improves customer trust and assists in monetization by permitting third party and analytics designers to monetize their services. Intermediate resource provisioning is adopted for an efficient decision on load partitioning. For provisioning of resources, a deep learning model is executed on various resource configurations having different combinations of Control Unit, GPU, and RAM. They are profiled and used for training data. Offload execution partition the deep learning analytics tasks and distribute them among the external devices connected to the local network. A simple capacity based partitioning scheme is utilized for optimal offload execution. The processing of each layer cannot be divided for parallel processing because the partitioning scheme used in the design is not very granular.
%Deep Learning is useful for three methods, classification, regression and encoding.

\begin{table*}[!ht]
    \centering
    \caption{Techniques that combines Machine learning and Edge Analytics}
    \begin{tabular}{|p{1.5cm}|p{6cm}|p{7cm}|}
      \hline \centering
         \textbf{Techniques} & \hspace{1cm} \textbf{Features} & \hspace{1cm} \textbf{Remarks}  \\ \hline \hline
        
ECNN \cite{song} & \small{$\bullet$ Enhances the performance of data aggregation and analytics \newline $\bullet$ Features or local inference are small in size and gives less transmission cost} & \small{$\bullet$ More data-intensive inference are generated at Cloud \newline $\bullet$ Edge nodes are unable to accommodate all layers of CNN} \\ \hline

Sarabia-Jacome \textit{et al.} \cite{Sarabia} & \small{$\bullet$ Container-based virtualization manages the limited resources \newline $\bullet$ Container-based virtualization technique uses the resources in a homogeneous manner \newline $\bullet$ Interoperable \newline $\bullet$ Secure \newline $\bullet$ Maintains privacy} & \small{$\bullet$ Deep learning models consumes high resources \newline $\bullet$ Training of deep learning models are performed in Cloud \newline $\bullet$ Generation of inference using deep neural networks takes more time \newline $\bullet$ Increase in connected Edge devices increases the inference generation time} \\ \hline

Moon \textit{et al.} \cite{Moon} & \small{$\bullet$ Edge node maintains data privacy \newline $\bullet$ Edge nodes control the actuators directly \newline $\bullet$ Cloud reduces computation in Edge nodes} & \small{$\bullet$ Framework more depended on Cloud \newline $\bullet$ Cloud performs data preprocessing \newline $\bullet$ Machine learning algorithm is executed in Cloud \newline $\bullet$ Models in the Edge analytics become obsolete after some time \newline $\bullet$ Obsolete models give incorrect decisions} \\ \hline

Dey and Mukherjee \cite{dl} & \small{$\bullet$ Standardization of resources gives low latency \newline $\bullet$ Standardization of resources enhances data manageability \newline $\bullet$ Intermediate resources provisioning provides an efficient decision on load partitioning} & \small{$\bullet$ Partitioning of Edge analytics task is very granular \newline $\bullet$ Processing of each layer cannot be performed in parallel \newline $\bullet$ Tuning of the hyper parameters of the DCNN requires more resources} \\ \hline
       
\end{tabular}
\label{ML_tab}
\end{table*}
% \end{longtable}

\section{Edge Analytic as a solution}
\label{sol}
Edge analytics is implemented in many fields such as Energy, Transportation \& logistics, Manufacturing, Retail, and many others. Some examples where Edge analytics is used are remote monitoring and maintenance for energy operations, monitoring of manufacturing \& logistics equipment, retail customer behavior analysis, and fraud detection in financial locations (ATMs) \cite{kdnuggets}. In this section, the Edge analytic solving issues present in some of the above fields are discussed. 

\subsection{Edge Analytics in Industry}
In this section, some Edge analytics are discussed, which are developed by companies for practical application.

Oracle company developed Edge analytics called Oracle Edge Analytics (OEA) \cite{OEA}. It aims to apply real-time intelligence to embedded devices for filtering, correlating, and processing events in real-time. OEA implements Oracle Real-Time Streaming Analytics technology to reduce memory usage. OEA is appropriate for industrial automation because it provides local storage and analysis, high-speed data collection and analysis, filtering, correlation, and pattern matching. For the management of appliances, OEA monitors events that may lead to downtime. In transportation and telemetry, OEA monitors various parameters such as intrusion detection, location tracking, temperature, and pressure. In healthcare, OEA performs data evaluation, analysis, automatic alerts, etc. OEA is enabled to integrate with lightweight adapters for events like JMS messages and sockets. It also provides the facility to easily use the adapter framework. OEA supports continuous query execution in event streams and persisted data stores. OEA implements ANSI SQL syntax for query operation accuracy. 
 
Siemens company developed Edge analytics called MindConnect \cite{MindConnect}. MindConnect monitors continuously the condition of the machine executing MindConnect. It determines any anomalies and helps to prevent unplanned downtime. It uses an analytics algorithm for preprocessing highly repetitive data. It filters the data that required to be sent to MindSphere. MindSphere \cite{MindSphere} is an operating system that helps in connecting, collecting, and retrieving context from IoT data. MindConnect performs fast machine monitoring and data visualization. It reduces maintenance costs by performing predictive maintenance and root-cause analysis. In industries, MindConnect monitors the vibration of the machines. It analyzes some parameters for the evaluation of vibrations. The upper limit of vibration is defined in MindConnect. It raises the alarm in case the vibration crosses the defined upper limit value. 

Accenture Labs have developed an Edge analytics framework \cite{accenture1} using a knowledge-based approach. The framework is developed for the heterogeneous environment, which is capable of implementing various applications, hardware, and model infrastructure. The framework has a hierarchical architecture that combines Edge devices, fog, and Cloud. These layers provide computation, storage, and network with different complexity and latency appropriate for implementing Edge analytics. Edge computing significantly reduces the latency, which boosts up the communication speed \cite{LA20193}. The framework has separated various components for easy replacement with new versions. The layers communicate with each other by passing asynchronous messages to connect with sensors or communicate with various components using open libraries. Open libraries help to extend the framework to support any business-specific or industry-specific custom protocols. The framework uses containerization technology to provide an abstraction over the complex connected Edge devices. Containerization supports a standardized deployment environment. The developers can easily build applications for Edge analytics. The abstraction helps with portability. In addition, it helps in the deployment of various Edge applications and models on various Edge computing hardware, ignoring device-specific capabilities, configurations, and settings. Likewise, in Cloud, an asynchronous event-hub abstracts the complex connected Edge devices. In the Fog layer, the abstract also has a cloud-based knowledge graph and an intelligent server. The cloud-based knowledge graph stores metadata regarding the hardware capability of Edge devices, the data format of sensors, and protocols. It also helps to find the Edge devices that have a specific Edge analytics application. When an Edge device requires some resources, the intelligent server validates the resource requirements using a knowledge graph and then provides the resources.
%It is a single interface for communication between Edge applications and sensors, other Edge applications and other components of the Cloud. 

Cisco \cite{Cisco1} proposed an Edge Analytics Fabric System. This system is a new approach to implement hyper distribution to increase the system scope, geography, and topologies. The system is modular to make the component independent and decoupled from other components. Moreover, it permits the addition and deletion of a new component in the system. The Edge component performs the data aggregation, filtering, and compression to reduce network cost, optimize the performance and increase scalability.  Microservices are software modules in Edge nodes that provide one or more functions but may not be a complete application. The advantages of microservices are (a) Filters and reduce the data that need to be transferred to an on-premises data center or Cloud.  (b) In case the client of the data is local and requests a quick response (e.g., factory control-loop feedback system), the data is analyzed in the Edge analytics, and (c) Captures the time series of the occurrence of an event. Data are collected, time-stamped, and stored in the Edge node in a historian database. Microservices executes in Edge nodes, fog nodes, data center, and Cloud. Cisco’s microservices are very flexible because it does not have the size or structural requirement except it requires an interface to message the router system. Communication between microservices is loosely coupled (i.e., asynchronous) to have unlimited scalability. For communication, it uses a message router that is connected to all microservice modules. The single message router is capable of handling hundreds of thousands of messages per second. The framework also has a data transport subsystem that is robust against microservice failure and message router failure. The scalability is further increased by adding more message routers. Edge nodes also have a specialized database. The advantages of the database are (a) data may not be required to send to Cloud, for example, local closed-loop supervisory system (b) data cleaning before transmitting to Cloud, and (c) the data is time-stamped in case of time series analysis.
%Message router sends the message to receiver microservice module. 

Xu \textit{et al.} \cite{Xu} designed a pioneering service called Edge analytics as a service (EAaaS) for IoT systems. The EAaaS service manages the real-time data stream on Edge analytics on IBM Watson IoT Platform. The service is more scalable, has low latency, and lightweight. The RESTful interfaces and dashboards is generated by the service for monitoring of status and analytics management. The service provides a unified rule-based analytic model. The model describes the rule condition, transformation, and user-defined data collection. The possible actions are decided using the model. But, the model is incapable of providing cross-device data Edge analytic. The service is also implemented as an Edge analytics engine which is lightweight and single threaded. The engine serves as the analytics runtime on the Edge gateways. However, the Edge analytics engine does not have software upgrade capability. 
%for real-time data streaming. With both low consumption of resources of gateways ( such as memory and CPU) and low latency, the engine is able to handle data streams of the incoming devices. 

\subsection{Retail}
Retail stores use many Edge devices such as CCTV, sensors, and Wi-Fi networks to collect raw data. For example, the number of customers with respect to time of the day, coupons used, images, and videos. The raw data is analyzed to predict customer behavior. The customer behavior helps to provide offers to attract customers to increase purchases from the shop. However, if these data are sent to Cloud for analysis, getting the output data will be very late. Hence, quick analysis is required for attracting the customer before they leave the shop. In such scenarios, Edge analytics is beneficial. 

%retail
Soni \textit{et al.} \cite{soni} proposed an Edge analytics service-oriented framework for personalized and real-time recommendation. The proposed framework is based on the interaction screen. The customer's purchase intents are collected from the social media activities. Edge analytics framework employs analytics uniformly across the channels to provide an efficient and scalable manager for the huge number of customers. The Edge information and social media profile data are matched based on the common attributes of the customers, such as name and address, using matching algorithms. Edge analytics only considers recent transaction details and a few social information of a customer. It helps to deliver personalized recommendations to a customer in real-time with low-cost data management. However, correctly linking the two entities, i.e., the social media entity and the enterprise entity is difficult. Because a single customer, i.e., an entity may have multiple accounts with different or similar names in social media and multiple accounts with the bank. The advantages and remarks of the framework are listed in table \ref{REA_tab}. 

\subsection{Agriculture} 
Smart technology is explored to improve agriculture, for example, installing sensors to keep surveillance on animals entering the crop fields, periodic watering, weather warning, etc. Nonetheless, constant human intervention is required, for example, sending a warning to the owner to remove animals from the crop field. Another solution is exploring Edge analytics. Video Edge analytic can maintain surveillance on animals. In case they come close to the crop fields, the sound is produced to scare them, as well as the owner, is informed. In case the farm animals graze outside the designated area, or some other animal/illegitimate person entered the area, the information is notified to the owner. Weather update is very important to maintain the quality of the crop. Suppose there is a possibility of rainfall in the afternoon. Then in the morning, watering of crops is avoided. Similarly, if a day is scorching, then the number of times of watering of crops increases. 

Bhargava \textit{et al.} \cite{dairy} proposed an efficient technique for precision dairy farming by using Edge analytics. The data are collected from three different Edge devices, namely, infield sensor nodes, collar devices, and Edge gateways. The infield nodes observe the farm environment and weather conditions. Collar devices are used to monitor the dairy cow's health and movement. The data from these devices are sent to the Edge gateways and later sent to the Cloud. The technique employs a delay-tolerant application since there is usually poor Internet connectivity in the farm environment. This leads to the easy transfer of data from Edge gateways to the Cloud. GPS data is observed to know the cow's broad location. The GPS sampling frequency is reduced to once per 15 minutes due to memory constraints. This arrangement increases the average lifetime of the device and also reduces energy consumption. The system compresses the raw data on the collar devices locally. Linear-SIP is used for data compression to improve the storage space and reduce the operational time. The Edge analytic helps in maintaining surveillance on the cows without any human intervention. The advantages and remarks of this technique are listed in table \ref{REA_tab}.

\subsection{Transportation}
Intelligent transportation systems (ITS) is a service provided by the smart city \cite{ITS}. The aim of ITS is to provide vehicles and transportation infrastructure along with sensing, connectivity, and autonomy. Moreover, it provides fast and safe travel. It is achieved by embedding the vehicle with smart sensors, i.e., Edge devices that collect many heterogeneous data regarding the passenger, it's surrounding, the vehicle itself, and other vehicles within close proximity. These data need to be collected at high speed and in real-time. In addition, for smart transportation, lots of decisions the vehicle itself has to make rather than the driver. However, sending these data to Cloud, computing the data, and sending back the result to take action drastically reduces the QoS. Moreover, delays in making decisions may be disastrous such as congestion and accidents. Therefore, ITS requires an efficient Edge analytics, which helps in taking timely action based on real time data while maintaining time constraints. 

Various sensors are embedded in the vehicle that measures vehicle dynamics, fuel consumption, environmental features, and driver fatigue level \cite{FAOUZI}. These are real-time data and have a huge volume. Thus, removal of unwanted data is essential because Edge devices have small storage space. Moreover, an overflow of data may lead to losing important information. After filtering, the data that need to be processed by Edge analytics is still huge. Thus, Edge analytics has to be efficient to quickly compute such a huge volume of real-time data. Moreover, QoS of ITS is more when the Edge device will compute quickly to determine optimal path. A self-driving vehicle has to automatically determine the direction, speed, and acceleration. However, these factors are affected by the surrounding of the vehicles, passengers, and traffic lights. In addition, traffic congestion plays an important role in determining the traveling path. Sometimes unprecedented situations (e.g., accidents) block or increase congestion on the road. To avoid taking such a road, the real-time data needs to be computed, and path change needs to be suggested to the driver. In the vehicles, both self-driving and driver controlled driving features are available. These vehicles can quickly change the modes. In driver controlled mode, the vehicle constantly provides suggestions such as a traveling plan, a vehicle within dangerous distance with other vehicles, etc. However, giving such suggestions is difficult due to unexpected decisions taken by the human driver. Edge analytics need to be robust because in case an attacker manipulates the computing of Edge analytics and it gives wrong decisions; then the consequences may be dangerous. In self-driving mode, it is extremely dangerous because the human driver may not be able to control the vehicle on time. Another important point to consider is that Edge devices also need to be secure because incorrect data also results in wrong decisions taken by Edge analytics.

Jiang \textit{et al.} \cite{Jiang} proposed a blockchain integrated video Edge analytics for ITS. It consists of two Edge devices, namely, vehicle and roadside unit (RSU) node. The camera is located in the vehicle node that performs video related tasks, and the roadside unit has the Edge analytics. The vehicle performs the partition of video into sets of frames and compresses them, and then those are forwarded to Edge analytics using vehicular networks. Blockchain \cite{Mollah} helps in secure data sharing among various vehicles and video storage in the RSU node. The video Edge analytics process is partitioned into various modules due to limited resources. The modules are decoding, pre-processing, object detection, object tracking, semantic segmentation, stereo, and optical flow. The video decoding module decodes/decompresses the encoded video and in real-time. The real-time streaming protocols decode/decompress the stream of video into sets of frames. The pre-processing module corrects non-uniform lights/colors and adjusts the intensity. The object detection module detects semantic object instances which belong to a certain class. It is important for achieving autonomous driving. Deep learning techniques are implemented for object detection. The object tracking module determines the location of an object or many objects after various time intervals. Also, it is used for determining traffic on the path taken by the vehicle. The semantic segmentation module helps in understanding the environment of the path, which helps in determining a change in the direction of driving. It assigns a label to each pixel present in an image where labels belong to a predefined set of categories. The stereo module extracts three-dimensional (3D) information from 2D images of the camera. The optical flow module provides information to many tasks and also performs motion estimation and tracking. The proposed Edge analytics converts the video offloading and resource allocation problem into a discrete Markov decision process to maximize the reward. The discrete Markov decision problem is solved by using deep reinforcement learning methods. Some features and remarks of this technique are highlighted in Table \ref{REA_tab}.

\begin{table*}[!ht]
    \centering
    \caption{Edge Analytics techniques in Retail, Agriculture and Transportation}
    \begin{tabular}{|p{1.5cm}|p{6.5cm}|p{6.5cm}|}
      \hline 
         \textbf{Techni-ques} & \centering \textbf{Features} & \hspace{1cm}\textbf{Remarks}  \\ \hline \hline
         
Soni \textit{et al.} \cite{soni} & \small{$\bullet$ Independent from big main system \newline $\bullet$ Scalable \newline $\bullet$ Easy to manage or configure \newline $\bullet$ Computation depend on the recent transactions \newline $\bullet$ Low cost for managing data \newline $\bullet$ Provide personalized recommendations in real time easily and quickly} & \small{$\bullet$ Difficult to link social media and the enterprise entity correctly \newline $\bullet$ Difficult to find similar entity using social media accounts} \\ \hline

Bhargava \textit{et al.} \cite{dairy} & \small{$\bullet$ Reduces energy consumption \newline $\bullet$ Compresses the raw data on the collar devices locally \newline $\bullet$ Compression reduces the memory consumption \newline $\bullet$ Low operational time \newline $\bullet$ Use of L-SIP does not lead to information loss} & \small{$\bullet$ Compression performance changes with change in variation of signal \newline $\bullet$ Tradeoff is maintained between the RMSE and the memory gain of L-SIP} \\ \hline
    
Jiang \textit{et al.} \cite{Jiang} & \small{$\bullet$ Smart contracts helps to secure data sharing and storage \newline $\bullet$ Video offloading and resource allocation problem is formulated as a Markov decision process to optimise transaction throughput and reduces latency \newline $\bullet$ Decentralized control with low complexity} & \small{$\bullet$ Blockchain makes trade-off between security and decentralization \newline $\bullet$ Convergence performance is bad when learning rate is small \newline $\bullet$ Average reward decreases with the increasing average size of transaction \newline $\bullet$ Average reward decreases with increase in maximum block interval} \\ \hline 
    
  \end{tabular}
    \label{REA_tab}
\end{table*}
% \end{longtable}

\subsection{Industry} 
The manufacturing industry highly requires Edge analytics; for example, in an offshore oil refinery, the oil rig has a large number of sensors (in thousands). The sensors continuously measure many parameters such as pressure, temperature, and gas emission. However, the majority of data produced by the sensors are not useful. Moreover, sometimes critical actions need to be taken quickly without enough time to send the data to the Cloud and receive the response. In such cases, Edge analytics is more practical to perform the analysis. The Edge analytic process the data quickly and decides for action appropriate for that situation. An Edge gateway is implemented in IoT by ELM Energy \cite{Nakod}. ELM Energy provides the solution for mining company clients. The client operates an off-the-grid mine, and this mine manages many electrical devices (e.g., solar battery, generator). The mine wants real-time power source management and less power wastage; for example, when the power supply decreases during cloudy or rainy days, the system should decide and use battery power or increase generator output. The ELM Energy solution performs processing in Edge analytics and gives real-time responses for many decision-making actions. The pre-processed data is later sent to Cloud. 

Edge analytics is used in upstream automation for artificial lift-assisted production \cite{saghir}. The Edge analytics implements machine learning algorithms in well-head. The operating system in Edge analytics is a general-purpose operating system (GPOS). GPOS is lightweight and executes efficiently along with the constraints of Edge analytics. Moreover, it manages the resources for the efficient execution of machine learning algorithms. Edge analytics have improved memory storage, enabling data management and processing of machine learning models (MLM). The Edge gateway deployed in Supervisory Control and Data Acquisition (SCADA) framework \cite{scada} requires an interface with an existing Remote Telemetry Units (RTU). Edge gateway only forwards the inference feedback to RTU and runs independently. When the Edge gateway sends new data to Cloud, MLM is updated based on the new data. Artificial lift mechanisms are used to improve hydrocarbon production and rod pumps in mature wells. Performance of rod pumps is determined by using dynagraph cards (i.e., position vs. load graphs). A dynagraph card is produced within a few seconds. Therefore, the storage of these cards becomes an issue. Therefore, it is handled by data file management and archiving locally. The data are collected by RTU and sent to the SCADA host. Determining any mechanical or production issues by an expert is time-consuming. Also, important information is lost when not analyzed within a certain time frame. Therefore, these tasks are performed using machine learning algorithms. A supervised learning algorithm is executed for training for the Cloud. Total 6000 historical dynagraph cards are used for training to predict any issue in production. The cards are pre-labeled, and classified by an operator or subject matter expert of rod pump operations. The historical dynagraph cards are translated into pixelated images. These images are used for training a generic model using a variety of machine learning algorithms. Prediction of MML is aggregated using the Ensembling technique, which adds more weight to the model based on accuracy achieved for a given problem class. The training of the model is repeated based on the required accuracy. After achieving the required accuracy, the MLM is sent to Edge gateways. These inferences require fewer resources for execution and can predict within seconds. 

%Industry
Boguslawski \textit{et al.} \cite{boguslawski} proposed an Edge analytics solution for monitoring of rod pumps. Edge analytics implements machine learning techniques. Edge analytics uses generalization or stacking ensembling techniques. The ensembling technique trains a meta-model using the trained models. The training of data is performed in the Edge devices. The meta-model helps in determining any failure of a process or machine. The inferences generated require fewer hardware resources. The Edge devices periodically save the data on the hard drive. The Edge analytics also perform in the absence of Internet connectivity. Disconnectivity from Cloud may lead to less accurate inferences because the training is performed upon small data size. When the connection with Cloud is established, the training model and the generated inferences are transferred to the Cloud. This technique can be deployed when the user wants to maintain data privacy. Ensembling techniques are highly computational, and Edge devices are unable to execute high computation machine learning algorithms. Therefore, implementing ensembling techniques that combine many machine learning algorithms is beyond the capability of Edge devices.

Kartakis \textit{et al.} \cite{Kartakis} proposed an Edge analytics for an end-to-end water leak localization and burst detection scheme. Edge analytics drastically reduces the communications cost compared to the traditional situations of periodical reporting between back-end servers and the sensor devices. Edge analytics utilizes the difference in the arrival times of the variations measured at sensor locations to localize water burst events effectively. For a particular scenario, the localization and Edge anomaly detection can find the anomaly within a 0.5 meter error due to the same time difference information produced by any burst event. The timestamps of events are sent to the back end. Reconfiguration of the pipe network is determined in the decision control process by using remotely controllable valves. For event detection, the timestamp of the data is given as input to Edge analytics. The data from the pairs of sensor nodes is examined recursively by Edge analytics for detecting anomalies. Moreover, Edge analytics perform local control functions with minimal latency that means the event detection and the early transient can also perform in sensor nodes. Edge analytics take compressed data as input, but the compression algorithms only compress small-sized data. The working space of the compression algorithms is limited to 10K. Moreover, the rate of data fluctuation rates is varied and based on the position of the Edge node in the topology. Furthermore, the performance of the compression algorithm affects the rate of data fluctuation.

\begin{table*}[!ht]
    \centering
    \caption{Edge Analytics techniques in Industry}
    \begin{tabular}{|p{2cm}|p{7cm}|p{5.5cm}|}
      \hline 
         \textbf{Techni-ques} & \centering \textbf{Features} & \textbf{Remarks}  \\ \hline \hline
Saghir \textit{et al.} \cite{saghir} & $\bullet$ A general-purpose operating system is used in the Edge analytics \newline $\bullet$ GPOS is lightweight \newline $\bullet$ GPOS executes efficiently along with the constraints of the Edge analytics \newline $\bullet$ Edge analytics improves memory storage by data management \newline $\bullet$ Edge gateway only forward the inference feedback to RTU \newline $\bullet$ Inferences require less resources for execution \newline $\bullet$ Inferences predict new data within seconds & $\bullet$ Supervised learning algorithm is executed for training in Cloud \newline $\bullet$ Edge analytics is dependent on Cloud \newline $\bullet$  Dynagraph cards are generated very quickly and occupies large memory \\ \hline
     
Bogus- lawski \textit{et al.} \cite{boguslawski} & $\bullet$ Meta-model helps in determining the failure \newline $\bullet$ Inferences generated requires less hardware resources \newline $\bullet$ Edge analytics also performs in absence of Internet connectivity \newline $\bullet$ Technique can be deployed when the user wants to maintain data privacy & $\bullet$ Disconnectivity from Cloud may lead to less accurate inference \newline $\bullet$ Ensembling techniques is highly computational \\ \hline

Kartakis \textit{et al.} \cite{Kartakis} & $\bullet$ Combines both the anomaly detection and the lightweight compression \newline $\bullet$ Produces an accurate and timely localized result \newline $\bullet$ Reduces the communications cost \newline $\bullet$ Effectively localize water burst events & $\bullet$ Finds the position of the anomaly with error 0.5 meters. \newline $\bullet$ Compression algorithms compress a small sized data\newline $\bullet$ Compression algorithm can influence the data fluctuation rate \\ \hline

  \end{tabular}
    \label{IEA_tab}
\end{table*}
% \end{longtable}

\subsection{Healthcare}
The patient wears bio-sensor or wearable sensors to monitor their condition. In contrast, Edge analytics help in deciding the action that needs to be taken in case of any abnormal condition of the patient. The Edge devices measure physiological signals such as blood pressure, heart rate, blood oxygen saturation level, skin temperature, etc. Also, the Edge device uses short-range radio such as Bluetooth, ZigBee, etc., for data transmission to an Edge node. The Edge node stores the data and performs the analysis. Edge analytics continuously monitors physiological signals received from the patient to detect any abnormal symptoms. When any abnormal symptoms are detected, some predefined action is initiated, for example, calling the ambulance or informing the concerned doctor. In this section, the techniques where Edge analytics is used for healthcare are precisely elaborated, and table \ref{HEA_tab} provides the advantages and remarks of these techniques.  

Madukwe \textit{et al.} \cite{Madukwe} proposed a healthcare solution based on Edge analytics. The architecture has three layers, namely, the device, gateway, and Edge analytics. The architecture is implemented in the Kaa IoT platform \cite{Kaa}. The patient carries a Bio-sensor that collects the data of the patient and transfer to a mobile application. Software development kit (SDK) of the platform pre-process the data. Then the data are transmitted to the server for analysis. Some data analysis is performed in Edge analytics to reduce the noise present in the data and maintain the security of the data. After the data analysis, the response is sent back to the smartphone using the MQTT protocol. MQTT protocol provides stable communication in weaker networks also while maintaining low latency and small packet size. Moreover, it has a push notification feature. 

Sanabria-Russo \textit{et al.} \cite{Russo} proposed an Edge node architecture for monitoring patients using data Edge analytics. The architecture consists of a container orchestration engine and virtualization tools, which are open source and lightweight. Edge devices perform live monitoring and real-time sensitive data processing with low latency while maintaining data privacy. The data Edge analytics tool is called Multivariate Statistical Process Control. The Edge device collects and monitors some parameters of the patient. Those parameter values are stored in CacheDB, which is a memory in the Edge device. The upper or lower threshold value of the parameters is predefined, but it varies for different patients. Thus, monitoring of parameters needs to be performed carefully in each patient’s Edge device. Edge analytics continuously computes the data. If any parameter exceeds its threshold value, the Edge device raises the alarm to a healthcare professional. All this information is periodically stored in Cloud for external visualization. 

%healthcare
Cardiac Health Management System (CHMS) \cite{cardiac} is proposed to improve clinical efficiency using Edge analytics. CHMS is a data-driven technique. The Edge devices collect phonocardiogram (PCG), i.e., physiological signal. The Edge analytics detects the abnormal or anomalous in PCG cardiac signals uses a robust machine learning algorithm. The machine learning algorithm uses boosting process because it does not result in overfitting due to weak learners. The Edge analytics uses Adaboost \cite{adaboost} boosting process. However, it is an iterative process and requires more energy and resources. The trained data are used to predict any abnormality in the patient. If any exception is found, then the result is sent to the Edge device to notify the user. The Edge device performs de-risking of private data with the application differential privacy.

%healthcare
Manocha and Singh \cite{motor} proposed a novel motor movement recognition framework assisted by Edge and Cloud analytics. It monitors the patient's physical activities, and in case some physical inactivity is detected, then appropriate actions are taken. The patient's physical activities are monitored using wearable sensors worn by the patients. The Edge nodes present in the architecture provide effective and efficient data processing since they are enabled with a Graphical Processing Unit (GPU). The Edge layer used for data processing enables efficient communication capabilities and reduces network congestion. The framework consists of four stages: data acquisition, Edge analytics, Cloud analytics, and suggestion generator. The Edge analytics stage consists of five steps. First, data are collected, i.e., the physical data about the patient's activities. Then, data is uploaded to a nearby Edge node in the connection establishment step. In the third step, data analysis is performed using deep learning algorithms in Edge analytics. The deep learning algorithm used is CNN which takes the input data and determines the movement type. The fourth step is information transmission; after diagnosing any medical abnormality, the analyzed data are sent to the required people to take further action. The final step is a suggestion generator, i.e., to take care of the patients by generating suggestions to solve the problem. The computation cost and the recognition time are low. The deep learning model provides stability to the system, which detects dynamics from unbalanced data in the real-world use-case. 

%healthcare
Fadlullah \textit{et al.} \cite{delay1} proposed an IoT Edge analytics approach using deep learning for healthcare. The deep learning algorithm used in edge analytics is a deep Convolutional Neural Network (DCNN). The sensors, i.e., Edge devices, are located in individual user's environments. The approach consists of three phases, namely, data gathering, raining, and prediction. In the data gathering phase, data is collected at periodic intervals from user's smartphones. The data is stored in the form of a load matrix which is input to the DCNN model. In the training phase, DCNN is utilized to create stable weight matrices from the input data. In the prediction phase, The future data load is decided using the training phase's weight matrix. In addition, the future data load prediction is utilized by the access points to predict the accuracy. The prediction error is low. The execution time changes with the batch size. The loss rate and accuracy also are improved. With these advantages, the user premises' access point detects anomalous test cases efficiently. Thus, it provides almost real-time analytics.

\begin{table*}[!ht]
    \centering
    \caption{Edge Analytics for healthcare}
    \begin{tabular}{|p{1.5cm}|p{7cm}|p{6cm}|}
      \hline \centering
         \textbf{Techni-ques} & \centering \textbf{Features} & \hspace{1cm}\textbf{Remarks}  \\ \hline \hline

Madukwe \textit{et al.} \cite{Madukwe} & \small{$\bullet$ Kaa is a robust platform \newline $\bullet$ Kaa provided data consistency and security \newline $\bullet$ Analysis in Edge analytics reduces noise in data \newline $\bullet$ Kaa provides a failure-proof connectivity \newline $\bullet$ MQTT protocol provides a stable communication in weak networks} & \small{$\bullet$ Threshold values of parameters are predefined but it varies for different patients}\\ \hline
         
Sanabria-Russo \textit{et al.} \cite{Russo} & \small{$\bullet$ Edge nodes perform live monitoring and real-time sensitive data processing \newline $\bullet$ Low latency \newline $\bullet$ Maintains data privacy \newline $\bullet$ Applicable for geographically distributed Edge devices} & \small{$\bullet$ Cloud is used for external visualization \newline $\bullet$ Parameters are predefined but it varies for different patients} \\ \hline

CHMS \cite{cardiac} & \small{$\bullet$ High accuracy \newline  $\bullet$ Privacy protection using de-risking \newline $\bullet$ Flexible for frequent retraining \newline $\bullet$ Reliable Edge analytics \newline $\bullet$ Robust Edge analytics \newline $\bullet$ Boosting process does not result in overfitting} & \small{$\bullet$ Boosting process requires more energy and resources \newline $\bullet$ Machine learning algorithm used not mentioned} \\ \hline

Manocha and Singh \cite{motor} & \small{$\bullet$ Low latency \newline $\bullet$ Low energy consumption \newline $\bullet$ Low computation cost \newline $\bullet$ Low recognition time \newline $\bullet$ Deep learning detects temporal dynamics from unbalanced data in a real-world use-case} & \small{$\bullet$ CNN does not encode the position and orientation of the object into their predictions \newline $\bullet$ Losses all internal information about the pose and orientation \newline $\bullet$ Deep learning algorithm are compute-intensive} \\ \hline

Fadlullah \textit{et al.} \cite{delay1} & \small{$\bullet$ Prediction stage has low error rate \newline $\bullet$ Reduces the loss rate \newline $\bullet$ DCNN improves accuracy} & \small{$\bullet$ Execution time depend on the batch size \newline $\bullet$ DCNN does not encode the position and orientation of the object into their predictions \newline $\bullet$ Loses all internal information regarding the pose and orientation} \\ \hline
          
  \end{tabular}
    \label{HEA_tab}
\end{table*}
% \end{longtable}

%%%%%%%% Smart Technology

\section{Smart Technology and Edge Analytics}
\label{ST}
Smart technology has made itself smart or intelligent by including computation in small smart devices such as smartphones and sensors. Using smart technology in the future smart town, smart home, smart waste management, the smart grid can be made practical. But to achieve this goal, smart technology will heavily depend on Edge technology. Smart technology wants to automate the activities of the devices, which requires the collection of every data, computing them, and deciding the action that needs to be taken. The Edge devices are placed everywhere to collect the data continuously and efficiently to address the above-raised issues. Connecting smart devices to Edge devices helps in the easy instantiation, privacy preservation, relocation, and upgrading \cite{Porambage}. To analyze the data, the Edge device takes the help of Edge analytics. Edge analytics to support the smart technology are embedding machine learning algorithms. However, the data generated by the Edge devices have some issues \cite{Sharma} which is making the embedding of the algorithms difficult. The data generated by the smart devices are raw data. But, the raw data cannot be given as input to the machine learning algorithm because it reduces accuracy. Thus, those algorithms require structured training data, i.e., tagged, categorized, or sorted for efficiency and accuracy. Moreover, data generated by the smart home or smart devices are small-sized and incomplete. Hence, training the machine learning algorithms on small-sized datasets reduces the accuracy. In addition, to reduce resource consumption, each smart application maintains a different date format. Thus, it makes the design of Edge analytics difficult because, for each application, different Edge analytics need to be designed. 

% Smartwatch %data Edge analytics
Takiddeen and Zualkernan \cite{smartwatch} proposed a smartwatch framework involving Edge analytics to increase the battery life. The framework divides the computations between the Cloud and the Edge device (i.e., smartwatch). Edge analytics reduces power consumption by detecting the abnormalities locally. In case any abnormality is detected, the Edge analytics sends warnings to the end-server. Edge analytics minimize power consumption by reducing the volume of collected data. It is achieved by data reduction, data prediction, and data transformation. In data reduction, the data are compressed because sending compressed data requires less power compared to sending fewer data. In data prediction, instead of data, a data model is maintained to predict the future data. Moreover, with the change of data, the data model is updated. Finally, in data transformation, raw sensor data are converted into information. 

% smart home, smart grid
Chen \textit{et al.} \cite{chen2019} proposed a smart power meter architecture assisted by Edge analytics for a smart home. The smart home architecture has DSM, which is present in the smart grid. DSM makes the smart home environmental-friendly by suppressing carbon emissions and reduces electricity costs. The framework consists of three components: Electrical Energy Management System (EMS), Smart AIoT (AI across IoT), and Push Notification service to achieve DSM in smart homes. The EMS depends on Cloud analytics. Moreover, EMS is essential for a smart home because it optimizes electrical energy consumption by monitoring and controlling its use. Smart AIoT is the smart power meter handled by Edge analytics. Smart AIoT is embedded with AI models and Arduino-MCU based, which performs the Edge analytics. Finally, the push notification informs the customer to achieve DSM. It also provides the knowledge about the remaining useful time of the appliances beyond which it will not perform at their best or the remaining lifetime.

%smart home
Lin and Hsiu \cite{smarthome} proposed a novel smart home architecture supported by Demand Side Management (DSM). In the proposed architecture, an additional flexible edge sensing device is embedded called Home Gateway (HG), which enables home appliances to be remotely controlled and monitored. The HG behaves as the master of Edge computing capabilities. HG is implemented using an Advanced RISC Machine processor-based system. The framework also has a set of Home Edge-Sensing (HES) devices that are networked together using wireless communication \cite{Chen20}. The HES devices are multifunctional and flexible. This feature also reckons for the real-time responsiveness of the architecture. Edge analytics is employed in these IoT-oriented smart homes because they provide real-time responsiveness for actions requiring emergency or can be handled locally, unlike the Cloud. Edge analytics use AI and machine learning algorithms for data processing to achieve real-time responsiveness. Moreover, Edge analytics is used to monitor and control home appliances, and it also contributes to DSM. A smart meter is utilized to receive Demand Response Programs. 

% smart water metering system
Amaxilatis and Chatzigiannakis \cite{spark} proposed a smart city water metering system using Edge analytics. The architecture consists of four layers: Edge devices, network bridge, Edge analytics, and Cloud. Similarly, the Edge devices comprise three meters: off-the-shelf water consumption meters, pressure meters, and automatic valves. The network bridge has a processor to compute the data. The Edge devices send the information about water pressure and water consumption as encrypted messages to the network bridge. The Edge analytics also has LoRa gateways and LoRa servers. Edge analytics performs a decryption process on the data packets and decodes them according to its data format requirements. The fourth layer comprises Cloud services that provide necessary interfaces and APIs. The Edge analytics platform is divided into two segments, Edge-1 and Edge-2 levels. The Edge-1 level collects the data packets, detects the message source, and uploads these messages to the higher layers. Edge-1 has low computational power. The Edge-2 level is more computationally powerful, therefore, performs analytics on more data. Various services are provided by Edge-2 processing services, such as performing analysis on incoming data packets, signal quality analysis, storing decryption keys of the meter, telemetry data analysis, and storing the analysis in a local storage layer. It also provides the required synchronization between the data and the central Cloud architecture. The framework is secure and reliable.

% smart meter
Sirojan \textit{et al.} \cite{Sirojan} proposed an intelligent Edge analytics technique for load identification in the smart meters. The method improves the Non-Intrusive Load Monitoring (NILM) of the metering unit in the prime circuit panel. Edge analytics reduces the cost, increases the scalability, improves accuracy, and improves methodologies for NILM. The Edge analytics use the high or huge frequency transient features for load identification because it gives better precision and scalability simultaneously. During the feature extraction process, the Discrete Wavelet Transform (DWT) is utilized to separate the transient signal into small distinguishable frequency bands in the process. Then, a neural network algorithm is used for load identification. After the load identification process, only the results are sent to service providers because the captured raw data becomes obsolete. 

%smart grid
Lin \cite{Lin} proposed a power-meter prototype based on Artificial Embedded Arduino for Edge analytics in smart grid. Arduino is a microcontroller with easy-to-use hardware and software. The Arduino MEGA microcontroller board consists of a current transducer, WizNet W5100 hardwired TCP/IP embedded Ethernet shield, real-time clock, and SD card.  The current transducer measures the current in the device under observation. WizNet W5100 hardwired TCP/IP embedded Ethernet shield is used for Internet connectivity. A real-time clock is a chip to record the timestamp using network time protocol. The measurement of load instances is collected and stored in the SD card. The Edge analytics execute an AI algorithm. The AI algorithm is the Back Propagation Artificial Neural Network (BP-ANN), which is locally deployed in Edge analytics. BP-ANN model is trained in Cloud, and then the model is embedded into Edge analytics. The features used by BP-ANN are electrical features such as real power and transient turn-on power.
%Edge analytics has a lightweight AI equipped smart power meter based on Arduino.

% smart grid
Oyekanlu \textit{et al.} \cite{Oyekanlu} proposed an Edge analytics for the smart grid using a Gaussian distribution function. Edge analytics uses the Gaussian distribution function for anomaly detection because it takes real-time data and consumes low memory. The Edge analytics has a low-cost processor and requires less computing power. The Gaussian distribution function fits in Edge analytics. All these features increase the efficiency of Edge analytics. The Gaussian distribution function support applications such as pattern recognition, machine vision, anomaly detection, and statistical machine learning. 

\begin{table*}[!ht]
    \centering
    \caption{Smart Technology and Edge Analytics}
    \begin{tabular}{|p{1.5cm}|p{6.5cm}|p{6.5cm}|}
      \hline \centering
         \textbf{Techniques} & \centering \textbf{Features} & \hspace{1cm} \textbf{Remarks}  \\ \hline \hline
         
\small{Takiddeen and Zualkernan \cite{smartwatch}} & \small{$\bullet$ Applied to various IoT applications \newline $\bullet$ Low power consumption \newline $\bullet$ High accuracy} & \small{$\bullet$ Maintains trade-offs between level of computational offloading, networks and latency constraints \newline $\bullet$ Less data available for guidance} \\ \hline

Chen \textit{et al.} \cite{chen2019} & \small{$\bullet$ Optimizes electrical energy consumption by monitoring and controlling the use \newline $\bullet$ Push notification provides useful information regrading the application such as life time} & \small{$\bullet$ AI model considered is compute-intensive \newline $\bullet$ AI model considered cannot be supported by Edge analytics \newline $\bullet$ Edge analytics consumes more energy and increasing the energy consumption} \\ \hline
          
Lin and Hsiu \cite{smarthome} & \small{$\bullet$ Low energy consumption \newline $\bullet$ Reduces electric costs and carbon emission \newline $\bullet$ Provides home health care facility \newline $\bullet$ Real-time responsiveness for actions} & \small{$\bullet$ Current technology does not allow connecting all appliances to central system \newline $\bullet$ Not implemented in large scale} \\ \hline

Amaxilatis and Chatzigiannakis \cite{spark} & \small{$\bullet$ Secure \newline $\bullet$ Good response time \newline $\bullet$ Reliable \newline $\bullet$ Reduces data communication \newline $\bullet$ Reduced congestion and small data size packet} & \small{$\bullet$ Edge analysis takes long time compared to Cloud server \newline $\bullet$ Encryption and decryption process increases time complexity} \\ \hline

Sirojan \textit{et al.} \cite{Sirojan} & \small{$\bullet$ Improves NILM methodology \newline $\bullet$ Provides better accuracy \newline $\bullet$ High scalability \newline $\bullet$ Uses powerful and low cost commodity embedded hardware} & \small{$\bullet$ Raw data are not transferred to Cloud \newline $\bullet$ Neural network is compute-intensive \newline $\bullet$ In case of near-real time data, huge volume of data are transferred to Cloud \newline $\bullet$ Near-real time analysis requires a high sampled data \newline $\bullet$ Near-real time analysis makes the system slow} \\ \hline

Lin \cite{Lin} & \small{$\bullet$ Reduces carbon emissions and electricity cost \newline $\bullet$ Arduino is lightweight \newline $\bullet$ AI computation is not performed in Edge analytics, so saves resources} & \small{$\bullet$ Training of AI algorithm dependent on Cloud \newline $\bullet$ Trained model is not updated} \\ \hline

Oyekanlu \textit{et al.} \cite{Oyekanlu} & \small{$\bullet$ Gaussian distribution function takes real time data \newline $\bullet$ Edge analytics with gaussian distribution function are of low-cost \newline $\bullet$ Gaussian distribution function consumes less memory \newline $\bullet$ Edge analytics has a low-cost processor} & \small{$\bullet$ Amplitude of the close Gaussian function is limited to maximum voltage \newline $\bullet$ Gaussian function’s height and width may be varied easily \newline $\bullet$ Noise in input signal gives incorrect result} \\ \hline

    \end{tabular}
    \label{tab:my_label}
\end{table*}

\section{Discussion}
\label{Dis}
Edge analytics will create enormous research possibilities and opportunities for the research communities. However, Edge analytics has lots of issues, and these issues of both Edge devices and Edge computing are also affecting the Edge analytics. One main issue of Edge analytics is the size of the Edge devices. The small size of Edge devices is restricting Edge analytics in embedding large and sophisticated hardware components and applications. Big size gives more space to include a big battery for the power supply or more computing components. In contrast, small-sized Edge devices are another goal of the manufacturer of Edge devices to increase mobility. These devices do not have the high computing power and require a reduction of computing load. One solution to reduce computation on Edge devices is caching. Edge caching \cite{Yao} stores computed responses and prevent repetitive execution of Edge analytics on the same input data. Thus, Edge caching is another crucial aspect that is required to enhance the efficiency of Edge devices. The prominent issue of Edge Computing is the limited power supply. An uninterrupted power supply is required for collecting data. Edge computation can replace the Cloud services, and thus, collecting data is more important, for example, CCTV for surveillance. There are also adversaries. The attacker targets the battery or power supply of Edge analytics. Attackers forward illegitimate user requests to keep the Edge devices engaged in continuous computing to exhaust the energy. It leads to the shut down of the Edge devices. Thus, there is a requirement for Edge security for low computing devices with less complex algorithms which will be deployed to provide security. In such a scenario, Bloom Filter is an excellent choice. Bloom Filter \cite{Bloom} is a probabilistic membership checking data structure that requires very little memory. Bloom Filter is widely used in networking and network security algorithms such as DDoS \cite{DDoS}. Moreover, Bloom Filter is a deduplicating data structure; hence, it can help in data filtering and reducing memory consumption \cite{Patgiri_2018}. 

Edge analytics embeds AI or machine learning algorithms for efficient and accurate analysis. Moreover, many techniques preferred deep learning algorithms for Edge analytics. The architecture of many techniques consists of many different types of sensors, for example, smart homes and intelligent transportation systems \cite{ITS}. These different sensors generate heterogeneous data having high dimensions. Thus, deep learning algorithms are efficient in computing such data because it's very appropriate to identify and model complex features in a data set with high dimensions. Deep CNN efficiently extracts the features from images and correctly detects the object \cite{lecun}. 

Currently, researchers are exploring federated learning (FL) for Edge technology. Federated learning \cite{FL} is a machine learning technique that trains an algorithm in a distributed manner, i.e., multiple devices train the algorithm. Federated learning is ideal for Edge analytics which has a constrained environment. An Edge analytics request for the training of a machine learning algorithm called an FL server. The FL server selects some Edge devices. These Edge devices train the model with their local dataset. Periodically, the Edge devices transfer model updates (i.e., weight and other parameters) to FL server for aggregation. This process continues till the required accuracy is achieved. Federated learning is an excellent technique to use machine learning algorithms for data analytics; regardless, many issues need to be addressed. The Edge devices selected for federated learning may not have the same dataset. Hence, the dataset needs to be transferred to multiple Edge devices, which incur additional communication costs. Sometimes the trained model may have class imbalance which leads to a decrease the accuracy. In such cases, an additional process for rebalancing is included. All selected Edge devices perform data augmentation \cite{Wong} on the minor classes before the training. Another important issue is communication cost. Each training requires many communication rounds among participant Edge devices and FL servers. Furthermore, a high-dimensional dataset requires more communication rounds. The selection of participant Edge devices is also an issue because it requires a load balancing algorithm. After all, a staggering node delays the completion of the training. Thus, federated learning is a good approach that needs to be explored for Edge analytics to break free from the dependency on Cloud.

Edge technology is a significant stepping stone for the next mobile generation, i.e., 6G \cite{nayak}. 6G promises low latency, low data rate, reliability, worldwide connectivity, etc. All these promises will fall flat without the help of Edge devices. Edge devices are the next mobile generation network nodes that will be deployed everywhere to fulfill the promises of 6G \cite{ZHOU2020253}. Moreover, the Internet of things (IoT) is moving towards the Internet of Everything (IoE) \cite{nayak2}. Thus, solving the issues of Edge analytics has become essential for achieving future technologies.

\section{Conclusion}
\label{Con}
Edge analytics is opening enormous new opportunities and possibilities to the research world. Users of Edge devices are producing Big Data, and transmitting all these data to the Cloud is a wastage of resources such as bandwidth, network, and Cloud storage. The majority of data is raw data. Therefore, using resources to send raw data, which may be deleted in the Cloud after the analysis, underestimates the value of resources. Edge devices help filter the data using Edge analytics and send information rather than data to the Cloud. Edge analytics is not limited to only filtering, but it shares the responsibilities of the Cloud. Edge analytics is executing in Edge devices to process data and tries to respond to the users' requests. Nonetheless, Edge analytics is still in its early days, with lots of issues and challenges that need to be solved, which are presented in the article.

\bibliography{mybib}

\end{document}